\documentclass[12pt]{amsart}
\usepackage{a4wide,palatino,epsfig,latexsym,natbib,amsmath,amsthm,graphicx}

\begin{document}

\title{Epistasis and entropy}  

\begin{abstract}
Epistasis
is a key concept in the theory
of adaptation.
Indicators
of epistasis
are of  interest for
large system
where systematic
fitness measurements
may not be possible.
Some recent approaches 
depend on  information theory.
We show that
considering shared entropy 
for pairs of loci
can be misleading.
The reason
is  that shared entropy
does not imply
epistasis for the
pair. This observation
holds true  also
in the absence of 
higher order
epistasis.
We discuss a refined
approach for 
identifying pairwise
interactions using 
entropy.
\end{abstract}

\author{Kristina Crona}  
\maketitle


\section{introduction}
Epistasis  tends to be
prevalent for antimicrobial
drug resistance mutations.
Sign epistasis means 
that the sign of
the effect of a mutation, whether
good or bad,  depends on background \cite{wwc}.
Sign epistasis may be important
for treatment strategies,
both for antibiotic 
resistance 
and HIV drug resistance
\citep{gmc, djk, bes}.
For instance, there are
sometimes constraints
on the order in which 
resistance mutations
occur. A particular resistance
mutation may only be selected
for in the presence of
another resistance mutation.
It is important to identify
such constraints.
A first question is
how one can identify
pairwise epistasis in a large
system. We will discuss
entropy \citep{s} and epistasis.
Information theory has been used
for HIV drug resistance mutations
\citep{ga} and more extensively
for analyzing human genetic
disease \citep[e.g.][]{dcw, kyz, slo}.
For recent review articles on 
epistasis and fitness landscapes
see e.g. \cite{h, kk}, and for
an empirical perspective \citep{ssf}.

\begin{figure}
\begin{center}
\includegraphics[scale=0.6]{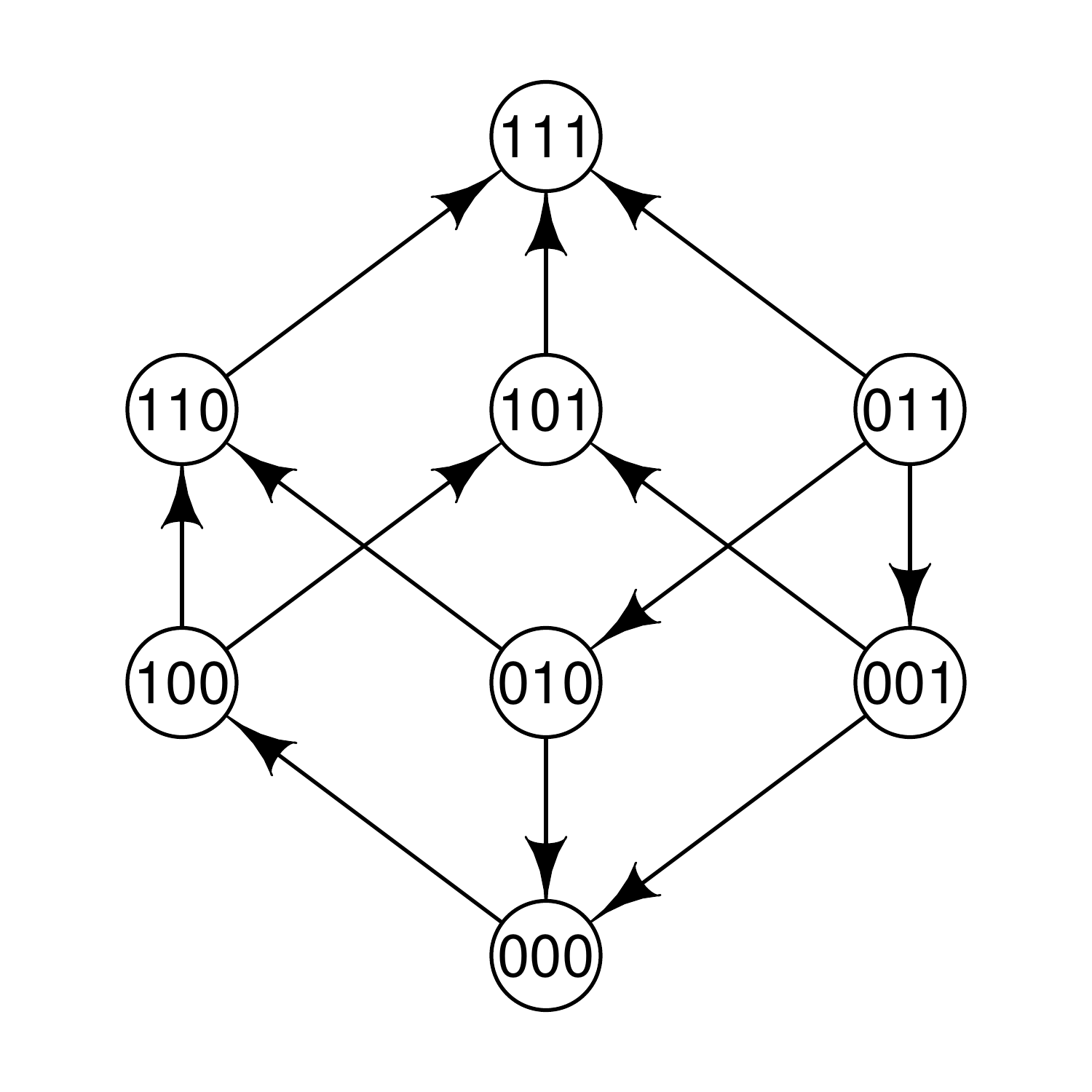}
\end{center}
\caption{systems}
\end{figure}

\section{results}
It is well established that 
genotypes are expected to
be in equilibrium proportions
if there is no epistasis in the
system, i.e., if fitness is multiplicative.
For instance, if two rare mutations
have frequencies $p$ and $q$,
then the frequency of the
genotype combining the
two mutations is expected to be
close to $pq$.
This statement holds true
regardless if recombination
occurs or not \citep{ol}.

We will  explore the
relation between entropy and
epistasis for a system with 
constraints as described
in the introduction.

Consider a 3-locus balletic system
where a mutation at the first locus confers resistance,
whereas mutations at the second and third loci
are only selected for in the presence
of the first mutation (otherwise  they are deleterious).
 We represent the case with a fitness graph \citep{cgb} (Figure 1).
As conventional, 000 denotes
the wild-type. 
For instance, one obtains a system with
the fitness graph as in Figure 1
for the log-fitness values

\begin{align*} 
&w_{000}= 0,   \quad w_{100}=0.09531018, \quad  w_{010}=-2.302585, \quad w_{001}= -2.302585,\\
&w_{110}= 0.1906204,  \quad  w_{101}=0.1906204, \quad w_{011}= -4.60517,   \quad w_{111}= 0.2859305
\end{align*}

The gene interactions for a 3-loci system 
can be described by the
sign pattern of 20 circuits,
or minimal dependence relations \citep{bps}.
The relevant two-way interactions
in this context be described by the six 
circuits corresponding to 
the faces of the 3-cube.
Specifically,

\begin{align*} 
&w_{000}-w_{010}-w_{100}+w_{110}>0\\ 
&w_{001}-w_{011}-w_{101}+w_{111}>0\\
&w_{000}-w_{001}-w_{100}+w_{101}>0\\
&w_{010}-w_{011}-w_{110}+w_{111}>0\\
&w_{000}-w_{001}-w_{010}+w_{011}=0\\
&w_{100}-w_{101}-w_{110}+w_{111}=0 \\
\end{align*}
The four inequalities 
express that there is positive epistasis
for the first and second loci, as
well as for the first and third loci.
The two equalities
show that there is no epistasis
for the second and third loci, regardless
of background.
The total 3-way epistasis is zero as well, 
\[
w_{111}-w_{110}-w_{101} -w_{011}+w_{100}+w_{010}+w_{001} - w_{000}=0. \\
\]

Higher order gene interactions have also been
described using Walsh coefficients \citep{wlw, pkr}.
For this landscape the Walsh coefficient $E_{011}=0$,
which indicates an absence of background averaged
epistasis for the second and third loci.

We will consider entropy during the process
of adaptation for this landscape.
 The starting point for adaptation
is the wild-type 000. We use a standard 
Wright-Fisher model
for an infinite population with
mutation rate $\mu=10^{-7}$.
The gene frequencies and
shared entropy after the given number of 
generations are listed in the table.

\begin{table}[ht]
\caption{Gene frequencies and shared entropy $I(2,3)$ for an infinite population with mutation rate $10^{-7}$.}
\centering
\begin{tabular}{c c c c c c c c c l}
\hline
\hline 
generations & 000 & 100 & 010 & 001 & 110 & 101 & 011 & 111  & I(2,3)\\
130& 0.7692  & 0.1850  & 0 & 0 & 0.0214& 0.0214& 0& 0.0031    &0.003206041\\
140& 0.4834 & 0.3015 & 0 & 0 & 0.0904 & 0.0904  &0 & 0.0343   &0.01736237 \\
146& 0.2723 & 0.3008, & 0 & 0 &    0.1597   &  0.1597 & 0 &   0.1075 &    0.02335234 \\
150& 0.1569 & 0.2539 & 0 & 0 & 0.1974 & 0.1974 & 0 & 0.1944  &0.0211462\\
160 & 0.0229 &0.0959 & 0 &0  &  0.1934 & 0.1934 & 0& 0.4943  &0.006950302\\
170 &   0.0020 &  0.0216 & 0& 0 &   0.1132 &    0.1132 &  0 &  0.7501  & 0.001270666 \\
\hline
\end{tabular}
\end{table}

The shared entropy for the second and third loci
differs from zero. However,
there is no 2-way epistasis for the pair of loci.

By extrapolation, consider an
analogous system for
$L$-loci. Then $L-1$ mutations 
are selected for
only if the first mutation 
has occurred, but there are
no other interactions.
One would get non-zero shared entropy 
for $L \choose 2$ pairs of 
loci, although there is
2-way epistasis for
$L-1$ pairs of loci only.

\subsection{A pair with no epistasis and maximal shared entropy}
The landscape
\begin{align*} 
&w_{000}= 0,   \quad w_{100}=0 \quad  w_{010}=-2.302585, \quad w_{001}= -2.302585,\\
&w_{110}= 0.09531018,  \quad  w_{101}=0.09531018 \quad w_{011}= -4.60517,   \quad w_{111}=0.1906204
\end{align*} 
is closely related to the previous example.
Indeed,
the two-way interactions can be described 
by the sign pattern
\begin{align*} 
w_{000}-w_{010}-w_{100}+w_{110}>0\\ 
w_{001}-w_{011}-w_{101}+w_{111}>0\\
w_{000}-w_{001}-w_{100}+w_{101}>0\\
w_{010}-w_{011}-w_{110}+w_{111}>0\\
w_{000}-w_{001}-w_{010}+w_{011}=0\\
w_{100}-w_{101}-w_{110}+w_{111}=0\\
\end{align*}
and the total 3-way epistasis is zero:
\[
w_{111}-w_{110}-w_{101} -w_{011}+w_{100}+w_{010}+w_{001} - w_{000}=0.
\]
Also in this case,
there is no epistasis for the second and third loci.
Mutations at the second and third loci are
selected for only in the presence of a mutation at the first locus.
However, this fitness landscape differs from the previous example
in that a mutation at the first locus is neutral for the wild-type.

Suppose  that 50 percent of hosts
start a new treatment with 000 viruses, and  50 percent start
with the 100 genotype.  That could be realistic, for instance if the 100 
genotype had some resistance to a previously used drug.
By assumption, eventually one would have about 50 percent 000 genotypes and
50 percent 111 genotype in the total population.
Then $I(2,3) =2$ although there is no epistasis for 
the second and third loci.
This example also points at a fundamental problem
relating pairwise epistasis and entropy.
At the time when we have 50 percent 
000 genotypes and 50 percent 111 genotypes,
obviously no method can reveal pairwise epistasis.

\subsection{A refined approach}
We will discuss a refined
approach for identifying 
pairwise epistasis.
Suppose that we have identified shared entropy
for a particular pair of loci $\{k,l \}$.
Let $S^{k,l}_1$ denote the set of
loci such that the shared entropy
\[
I(k:i) \neq 0 \text{ or }  I(l:i) \neq 0 . 
\]
Let $S^{k,l}_2$ denote the
set of loci with non-zero shared
entropy for some locus in $S_1$,
and so forth. Let 
$S^{k,l}= \bigcup S_i \setminus{ \{k,l \}}$.

Let $v$ denote one of the
$2^{|S|}$ possible states
for $S$, 
and consider the 
subsystem of genotypes
determined by $v$.
If the shared entropy $I^v (k:l)=0$
for all $v$, then there is
no indication of
of epistasis for $\{l,k \}$.

We can apply the
refined approach for
the second and third
loci in our example 
where $I(2,3)=2$.
Then
\[
S=\{ 1 \}, \quad I^{(0)}(2:3)=I^{(1)} (2:3)=0.
\]
Consequently, there is no indication of epistasis for
the second and third loci.

The described method could be useful
for identifying cases with 
shared entropy and no epistasis.
However,
it remains to explore to what extent
the method is useful in a more
general setting.

\section{discussion}
We have demonstrated that 
shared entropy for two loci
does not imply epistasis
for the pair. This observation
holds true also in the absence
of 3-way epistasis 
in a single environment.
Entropy based approaches
to epistasis are coarse.
We have discussed a refined
approach which filters out
some cases where shared entropy
depends on states at other loci.

There are obviously other
reasons for caution in 
interpretations of entropy
for drug resistance mutations.
Different drugs constitute
different environments.
Some resistance
mutations may be
correlated if
they are beneficial in the presence
of a particular drug, but 
not for other 
drugs. In such cases
entropy would not
not imply epistasis.

Our results show that
observations on
entropy and epistasis 
based on  2-locus
systems can be misleading
for general systems.
From a theoretical point of view,
a better understanding of large 
systems would be useful for handling 
drug resistance data.

\section{methods}
Let $x$ and $y$ be discrete
random variables with states $x_1, x_2$
and $y_1, y_2$.
Let $p_i$ denote the frequency of $x_i$,
and $p_{ij}$ the frequency for the
combination of $x_i$ and $y_j$.
The entropy \citep{s} 
$H(x)$ and the joint entropy
$H(x,y)$ are defined as
\begin{align*}
H(x)=&-p_{1} \text{ log}(p_{1})-(1-p_1) \text{ log}(1-p_1)\\
H(x,y)=&-p_{11} \text{ log} \, p_{11} - p_{12} \text{ log} \, (p_{12}) \\
&-p_{21} \text{ log} \, p_{21}  - p_{22} \text{ log} \, (p_{22})
\end{align*}

The shared entropy is the quantity
$ 
I(x:y)=H(x)+H(y)-H(x,y).
$

In general $I(x:y) \geq 0$, and
the shared entropy is a measure of 
dependence.

\end{document}